\documentclass[preprint,review]{elsarticle}

\usepackage{bm}
\usepackage{verbatim}
\usepackage{amsmath,amssymb}

\begin{document}

\title{Moment independent expansion for fourth-order corrections in lattice Boltzmann methods}

\author{Kyle Strand}
\ead{kyle.t.strand@ndsu.edu}
\address{Department of Physics, North Dakota State University, NDSU Dept 2755, PO Box 6050, Fargo ND 58108-6050, USA}

\date{\today}

\begin{abstract}
A expansion to fourth-order for lattice Boltzmann methods is presented. This expansion provides an easy model for finding fourth-order corrections to lattice Boltzmann methods for various physical systems. The fourth-order terms can give rise to improved results over traditional second-order lattice Boltzmann implementations. Although, this manuscript solely deals with fourth-order expansions, this expansion is easily extended to arbitrary order. We present examples of how this expansion is utilized and provide basic analysis to show how the fourth-order methods differ from lower order models for both diffusive systems and phase separating systems.
\end{abstract}

\maketitle

\section{Introduction}

Since its initial development in the late 1980s, lattice Boltzmann methods have been growing as a powerful tool in computational physics. Originally introduced for simulating hydrodynamic systems \cite{mcnamarazanetti,higuerajiminez,higuerasucci}, the lattice Boltzmann methods have been an active area of research which consistently is being improved and finding new applications in other fields of physics, namely phase separation \cite{wagner1,wagner2}, electrostatics \cite{electrostatics}, quantum mechanics \cite{qm}, diffusion \cite{wolf,wagnerstrand,shan}, and moisture transport through barrier coatings \cite{strandfeickert,feickertwagner}. As a consequence of the many developments in lattice Boltzmann methods, deeper analysis on the fundamental methods have been required for better understanding as to how the method treats any given system \cite{reza}.

In order to show that a specific lattice Boltzmann model is simulating the desired equations of motion, the hydrodynamic limit of the lattice Boltzmann equation is studied. In principle, this arises from an expansion of arbitrary order. In most implementations, an expansion up to second-order is sufficient. In certain computational situations, additional higher order terms can be utilized to better match the physical behavior of a system \cite{dubois}. Cases such as this require higher-order expansions and analysis to develop these corrections to the method and to better match the physical system \cite{dubois,Philippi,lazzara}. Previously, research on higher-order lattice Boltzmann models have been developed to model the Korteweg-de Vries equation (Yan and Zhang)\cite{yanzhangkdv}, general nonlinear partial differential equations (Lin-Jie et al.)\cite{linjie}, as well as phase transitions (Siebert et al.)\cite{siebert}. Traditionally, higher-order expansions of lattice Boltzmann models have utilized an \textit{amending function}, which is utilized as a correction to reduce discretization errors which may be present in lower-order models. The physical meaning of these amending functions is not currently well understood. Recently, Otomo et al. developed lattice Boltzmann models which introduced scaling parameters in the local equilibrium functions which do not require the use of an amending function\cite{otomo}. This model has been justified using both the Chapman-Enskog and Taylor expansion methods.

This manuscript presents a moment independent expansion for examining the lattice Boltzmann equation in the fourth-order with the intention of easily finding the equations of motion which govern a specific system. This manuscript presents this expansion using the traditional amending function. Although the physical meaning of the amending functions are not entirely clear, the expansion presented allows for a simple implementation of these traditional corrections to lattice Boltzmann models. In section \ref{sec:two}, we introduce the lattice Boltzmann method. Section \ref{sec:three} shows the moment independent expansion of the lattice Boltzmann equation up to fourth-order and examines the hydrodynamic limit from which the terms of higher order for the method arise. Section \ref{sec:four} then shows applications of the proposed expansion to various simple one dimensional system. We re-derive results previously derived by Strand et al. \cite{strandfeickert} using this new moment independent expansion and reanalyze the way the hydrodynamic limit treats temporal derivatives in the diffusive case. We then present two separate derivations for fourth order phase separating systems. The first utilizes a chemical potential model and the second employs the traditional diffusive lattice Boltzmann moments with the addition of external forcing. We present comparisons between the new fourth-order methods and the second order method. Although, we only present results for fourth order methods, this expansion can be easily generalized to arbitrary order, which we present in an appendix.

\section{The lattice Boltzmann method}\label{sec:two}

The lattice Boltzmann equation is a discrete form of the Boltzmann equation which is discretized in both space and time. The equation takes the form
\begin{equation}
f_i(\bm{x}+\bm{v}_i \Delta t, t+\Delta t) = f_i(\bm{x},t) + \Omega_i + F_i
\label{eq:LBE}
\end{equation}
where $\Delta t$ is a discrete time step, $\Omega_i$ is a specified collision operator, $F_i$ is a forcing term which allows for the inclusion of external conservative forces \cite{shanchen,Gross}, $\bm{v}_i$ is an element of a set of prescribed lattice velocities $\{\bm{v}_i\}$ and $i$ is indicates a specific element of the velocity set. In principle, lattice Boltzmann methods are sets of the discrete-velocity particle distribution functions $f(\bm{x},\bm{v}_i, t)$. These distribution functions represent the density of particles moving with a discrete velocity $\{\bm{v}_i\}$ at a position $\bm{x}$ and time $t$. The number of the distribution functions utilized in a LBM is equivalent to the number of elements of the set $\{\bm{v}_i\}$. In more practical language, the set $\{\bm{v}_i\}$ is a set of vectors which connect the points of a lattice in various ways. The vectors in $\{\bm{v}_i\}$ are used in tandem with the discrete-velocity particle distribution functions to determine the motion of the particles around the lattice. In principle, the collision operator $\Omega_i$ should not change the conserved quantities of a system.

The distribution functions can be used to find the macroscopic quantities of a system through weighted sums known as velocity moments of $f_i(\bm{x},t)$. For example, hydrodynamic systems have macroscopic quantities density, $\rho(\bm{x},t)$, and momentum, $\rho(\bm{x},t) \bm{u}(\bm{x},t)$, where $\bm{u}$ is the macroscopic flow velocity, to define 
\begin{align}
\rho(\bm{x},t) &= \sum_i f_i(\bm{x},t)\label{eq:hydrodensity}\\
\rho(\bm{x},t)\bm{u}(\bm{x},t) &= \sum_i \bm{v}_i f_i(\bm{x},t).\label{eq:hydromom}
\end{align} 
It is noted that the density in Eqn. (\ref{eq:hydrodensity}) is a scalar and the momentum in Eqn. (\ref{eq:hydromom}) is a vector. In the current manuscript, we are concerned with the most general representations. For this reason, we will rename these moments to moment independent scalars and vectors in the form
\begin{align}
S(\bm{x},t) &= \sum_i f_i(\bm{x},t) \label{eq:scalardensity}\\
j_\alpha(\bm{x},t) &= \sum_i \bm{v}_i f_i(\bm{x},t).\label{eq:scalarmom}
\end{align}
Later, when we introduce higher order moments, we will extend these methods to higher ranked tensors.

The collision operator $\Omega_i$ will not modify the conserved quantities of a system. This is represented by using discrete-velocity moments of the collision operator defined by
\begin{align}
\sum_i \Omega_i(\{f_j\}) &= 0\\
\sum_i v_{i\alpha} \Omega_i &= 0\\
\end{align}
for a system which conserves both mass and momentum. The collision operator can take many different forms. It is common that a multi-relaxation time (MRT) collision operator is employed. The MRT collision operator uses particle collisions to relax the distribution functions $f_i$ to a local equilibrium distribution $f_i^0$ with a characteristic relaxation time $\tau_i$. The index $i$ refers to the relaxation time of a specific mode. The MRT operator takes the form.
\begin{equation}
\Omega_i = \Lambda_{ij}(f_j^0 - f_j)
\end{equation}
where $\Lambda_{ij}$ is a collision matrix with eigenvalues given by the relaxation times \cite{kaehler,PREGinzburg}. 

There is a special case of the collision operator in which all of the relaxation times are equal \cite{Qian}. This is equivalent to writing a diagonal collision matrix such that
\begin{equation}
\Lambda_{ij} = \frac{1}{\tau}\delta_{ij}.
\end{equation}
Using this special diagonal collision matrix with equivalent relaxation times gives a simplified form of the MRT collision operator which is written
\begin{equation}
\Omega_i = \frac{1}{\tau}(f_i^0 - f_i) 
\end{equation}
This form of the collision operator is called the Bhatnagar, Gross, and Krook (BGK) collision operator. The equilibrium distribution is inherently a function of the macroscopic properties of the system. A discretized second order expansion of the Maxwell-Boltzmann distribution is commonly employed, but for the sake of the required generality, we will not require any specific definitions for the equilibrium distribution.

\section{General fourth-order expansion of the lattice Boltzmann method}\label{sec:three}
The equations of motion for any given system can be derived to arbitrary order from the lattice Boltzmann equation Eqn. (\ref{eq:LBE}). This can be achieved by taking a Taylor expansion of the left hand side of Eqn. (\ref{eq:LBE}) with $\Delta t$ being the small parameter. The Taylor expansion method for lattice Boltzmann methods is now well known in this form and was first introduced by Holdych et al. \cite{holdych1} and has been utilized in practice in other lattice Boltzmann analyses \cite{holdychtrunc,wagner3,duboistaylor}. At this point, for simplicity in the remaining derivation, we set $\Delta t = 1$ and we are left with a PDE in terms of the distribution functions $f_i$ in the form
\begin{align}
(\partial_t + v_{i\alpha}\partial_\alpha)f_i &+ \frac{1}{2}(\partial_t + v_{i\alpha}\partial_\alpha)^2f_i + \frac{1}{6}(\partial_t + v_{i\alpha}\partial_\alpha)^3f_i \nonumber\\
&+ \frac{1}{24}(\partial_t + v_{i\alpha}\partial_\alpha)^4f_i + O(\partial^5) = \frac{1}{\tau}(f_i^0 - f_i) + F_i\label{eq:fipde}
\end{align}
where we have used the BGK collision operator with an additional forcing term. The inclusion of the Greek indices indicate the Einstein summation convention. However, we desire a partial differential equation for $f_i^0$ for equilibrium behavior. In order to do this, we can rewrite Eqn. (\ref{eq:fipde}) such that 
\begin{equation}
f_i = f_i^0 - \tau\left[F_i + (\partial_t + v_{i\alpha}\partial_\alpha)f_i+\frac{1}{2}(\partial_t + v_{i\alpha}\partial_\alpha)^2f_i + \frac{1}{6}(\partial_t + v_{i\alpha}\partial_\alpha)^3f_i \right] + O(\partial^4).
\label{eq:fibgk}
\end{equation} 
With this form for $f_i$, we can iteratively substitute Eqn. (\ref{eq:fibgk}) into Eqn. (\ref{eq:fipde}). This process will then allow us to find a PDE for the equilibrium behavior. It should be noted that are able to write Eqn. (\ref{eq:fibgk}) up to third order since each subsequent iterative substitution of $f_i$ will be under additional derivatives.

After repeating this iterative process and rearranging terms \cite{wagner3}, we arrive at fourth-order partial differential equation for the equilibrium distribution and forcing terms in the form 
\begin{align}
(\partial_t + v_{i\alpha}\partial_\alpha)(f_i^0 &- \tau F_i) - \left(\tau - \frac{1}{2}\right)(\partial_t + v_{i\alpha}\partial_\alpha)^2(f_i^0 - \tau F_i) + \left(\tau^2 - \tau + \frac{1}{6}\right)(\partial_t + v_{i\alpha}\partial_\alpha)^3(f_i^0 - \tau F_i)\nonumber\\ 
&- \left(\tau^3 - \frac{3}{2}\tau^2 + \frac{7}{12}\tau - \frac{1}{24}\right)(\partial_t + v_{i\alpha}\partial_\alpha)^4(f_i^0 - \tau F_i) + O(\partial^5) = \frac{1}{\tau}(f_i^0 - f_i) + F_i. 
\label{eq:fi0pde}
\end{align}

We will now employ moments of the equilibrium distributions to derive general equations of motion in the hydrodynamic limit.

In Eqns. (\ref{eq:scalardensity}-\ref{eq:scalarmom}), we presented the moments of the discrete-velocity particle distribution functions which would reproduce the macroscopic properties of the system in terms of general scalars and vectors. Since we have derived a partial differential equation of motion for the equilibrium distribution, we can now extend these macroscopic moments for the $f_i$ distribution functions to the $f_i^0$ distribution functions. The number of moments required must be equal to the degree of order of our PDE for $f_i^0$ plus one. This is due to the fact that for a PDE of arbitrary order, we have a term which is written as $(\partial_t + v_{i\alpha}\partial_\alpha)^n$, where there will be one term which has $n$ powers of $\bm{v}_i$. For the case of Eqn. (\ref{eq:fi0pde}) we need five moments since this equation is written up to fourth order. Since we can derive this PDE to arbitrary order, this requirement will hold to any order. In the fourth-order expansion we derived, we define the moments
\begin{align}
\sum_i f_i^0 &= S\label{eq:genscale0}\\
\sum_i v_{i\alpha} f_i^0 &= j_\alpha\\
\sum_i v_{i\alpha}v_{i\beta} f_i^0 &= \Phi_{\alpha\beta}\\
\sum_i v_{i\alpha}v_{i\beta}v_{i\gamma} f_i^0 &= \Gamma_{\alpha\beta\gamma}\\
\sum_i v_{i\alpha}v_{i\beta}v_{i\gamma}v_{i\delta} f_i^0 &= \Xi_{\alpha\beta\gamma\delta}\label{eq:genscale4},
\end{align}
The zeroth order and first order moment produce a scalar and vector respectively as we had seen previously in the moments of the distribution functions. The additional second through fourth moments each give a tensor of rank which is equivalent to the number of velocities in each sum. These general tensors are then inserted into  Eqn. (\ref{eq:fi0pde}) when we sum over all $i$ in the hydrodynamic limit.

In Eqn. (\ref{eq:fi0pde}), we also have a dependence on the forcing terms $F_i$. These forcing terms also require their own distinct moments, but will have the same tensor ranking as seen in the moments for the equilibrium distribution. We define the moments to be
\begin{align}
\sum_i F_i &= 0\label{eq:force0}\\
\sum_i v_{i\alpha} F_i &= F_\alpha\\
\sum_i v_{i\alpha}v_{i\beta} F_i &= \Psi_{\alpha\beta}\\
\sum_i v_{i\alpha}v_{i\beta}v_{i\gamma} F_i &= \Delta_{\alpha\beta\gamma}\\
\sum_i v_{i\alpha}v_{i\beta}v_{i\gamma}v_{i\delta} F_i &= Z_{\alpha\beta\gamma\delta}.\label{eq:force4}
\end{align}
With the moments from both the equilibrium distribution and the forcing terms, we can then take the hydrodynamic limit of Eqn. (\ref{eq:fi0pde}) by summing over all $i$. 

Now summing over all $i$ in $\{\bm{v}_i\}$ in Eqn. (\ref{eq:fi0pde}) will then give a form for a moment independent fourth-order equation of motion in the defined tensor notation. After much rearranging, we are left with 
\begin{align}
  \partial_t S + \partial_\alpha j_\alpha &- \tau \partial_\alpha F_\alpha - \lambda_2(\tau)(\partial_t^2 S + \partial_\alpha\partial_\beta\Phi_{\alpha\beta} + 2\partial_t\partial_\alpha j_\alpha -2\tau\partial_t\partial_\alpha F_\alpha - \tau \partial_\alpha\partial_\beta \Psi_{\alpha\beta}) \nonumber\\
  &+ \lambda_3(\tau)(\partial_t^3S + 3\partial_t\partial_\alpha\partial_\beta\Phi_{\alpha\beta} + 3\partial_t^2\partial_\alpha j_\alpha + \partial_\alpha\partial_\beta\partial_\gamma \Gamma_{\alpha\beta\gamma}- 3\tau\partial_t^2\partial_\alpha F_\alpha \nonumber\\
  &- 3\tau\partial_t \partial_\alpha\partial_\beta \Psi_{\alpha\beta} - \tau\partial_\alpha\partial_\beta\partial_\gamma \Delta_{\alpha\beta\gamma}) - \lambda_4(\tau)(\partial_t^4 S + 6\partial_t^2 \partial_\alpha\partial_\beta \Phi_{\alpha\beta} + 4\partial_t^3 \partial_\alpha j_\alpha \nonumber\\
  &+ 4\partial_t\partial_\alpha\partial_\beta\partial_\gamma \Gamma_{\alpha\beta\gamma} +\partial_\alpha\partial_\beta\partial_\gamma\partial_\delta \Xi_{\alpha\beta\gamma\delta} - 4\tau\partial_t^3\partial_\alpha F_\alpha - 6\tau \partial_t^2\partial_\alpha\partial_\beta \Psi_{\alpha\beta} \nonumber\\
  &-4\tau\partial_t\partial_\alpha\partial_\beta\partial_\gamma \Delta_{\alpha\beta\gamma} - \tau \partial_\alpha\partial_\beta\partial_\gamma\partial_\delta Z_{\alpha\beta\gamma\delta})\nonumber\\
  &= \sum_i \Omega_i
  \label{eq:stencil}
\end{align}
where $\lambda_m(\tau)$ are Bernoulli polynomials in $\tau$ \cite{zhangkdv1} such that
\begin{align}
  \lambda_1(\tau) = 1\label{eq:bern1}\\
  \lambda_2(\tau) = \left(\tau - \frac{1}{2}\right)\\
  \lambda_3(\tau) = \left(\tau^2 - \tau + \frac{1}{6}\right)\\
  \lambda_4(\tau) = \left(\tau^3 - \frac{3}{2}\tau^2 + \frac{7}{12}\tau - \frac{1}{24}\right)\label{eq:bern4}
\end{align} 
This equation is a moment independent fourth-order equation of motion with the inclusion of general forcing term. To tailor this to a specific system, all that is needed is to define each of the tensors in a form which satisfies the system. In the following section, we will show examples of how this is employed. It is important to note that the expansion presented in Eqn. (\ref{eq:stencil}) will not recover the equations of motion for all conserved quantities. For a diffusive system which conserves mass, but not momentum, this single equation would be sufficient. However, if we want to model a hydrodynamic system, we would need equations of motion for both mass and momentum. In order to get the momentum equation of motion, all that is needed is to multiply both sides of Eqn. (\ref{eq:fi0pde}) by an additional $v_{i\alpha}$ term and the sum can be performed as shown previously. In a momentum conserving system, the additional equation of motion would require an additional moment for both the equilibrium distribution and the forcing terms. These additional moments would introduce another tensor which is of fifth-rank. This same process can be performed for up to an arbitrary number of conserved macroscopic quantities following the same prescription. For the sake of space, additional details of the previous derivations and extending to $n^{th}$ order are covered in appendix \ref{sec:append}. 

\section{Employing the moment independent expansion to diffusive and phase separating systems}\label{sec:four}
In order to show the usage and validity of the expansion provided in Eqn. (\ref{eq:stencil}), we utilize moment definitions designed to model the diffusion equation and Cahn-Hilliard equation. For simplicity, we will present a D1Q3 lattice Boltzmann model for each system. The D1Q3 representation simulates motion of particles in one spatial dimension with a set of three velocities such that
\begin{equation}
\{\bm{v}_i\} = \{0,1,-1\}.
\label{eq:d1q3}
\end{equation}
The vectors in the set from Eqn. (\ref{eq:d1q3}) mean that the particles are restricted to a rest state and motion in to the neighboring lattice space to the left and right. For brevity, both methods being presented lack the external forcing terms. To add forcing terms, one must simply define the tensors in Eqns. (\ref{eq:force0}-\ref{eq:force4}). 

\subsection{Diffusion equation}
\begin{figure}
\includegraphics[width=\columnwidth,clip=true]{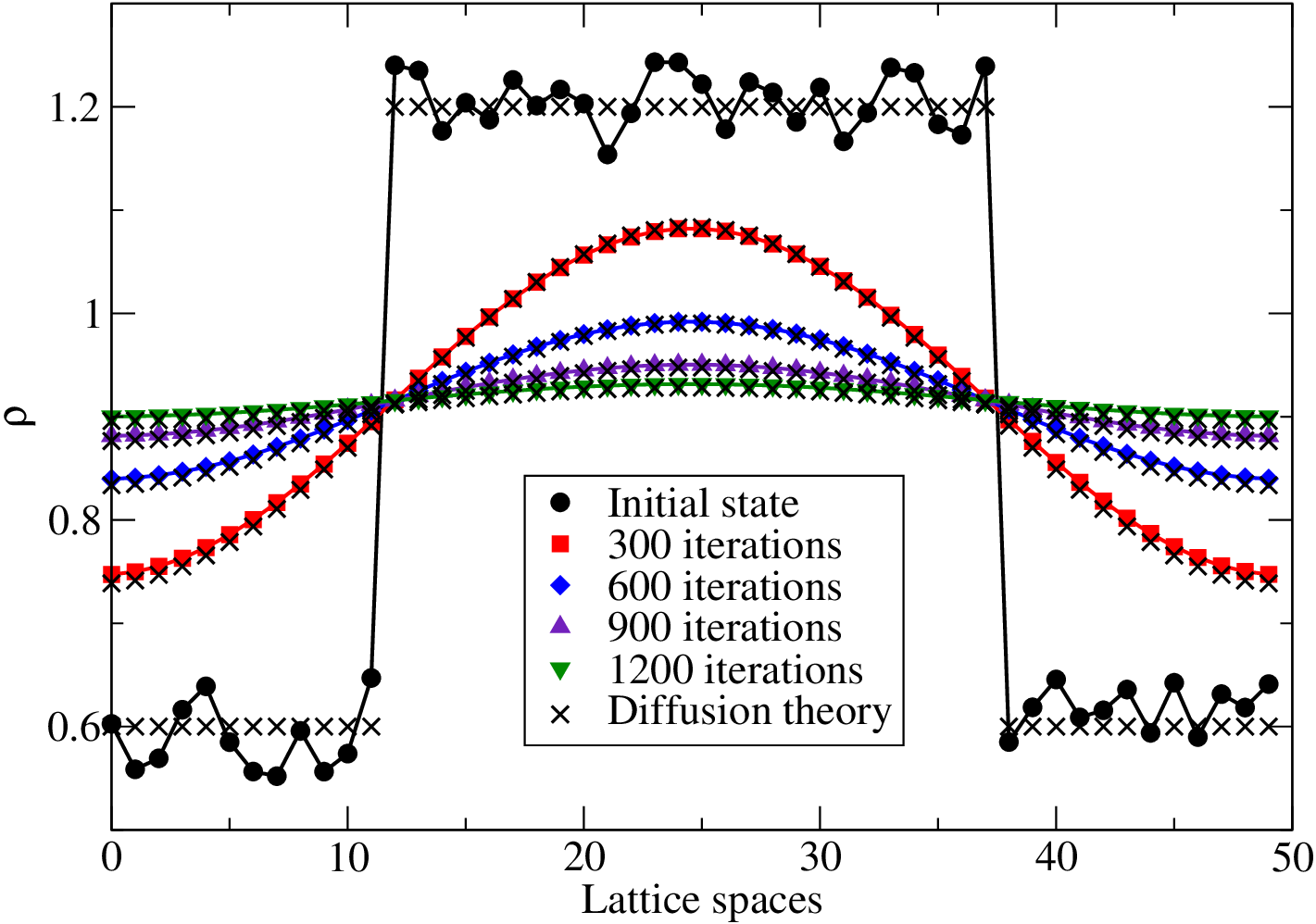}
\caption{Comparison of numerical solution of Eqn. (\ref{eq:telegraph}) (symbols), diffusion equation (X) and lattice Boltzmann simulations (solid lines) over various time evolved states for a diffusion front. The plot shows near perfect agreement between the numerical solution of Eqn. (\ref{eq:telegraph}) and the simulation results for all times. Slight differences between the central difference of the diffusion equation and the theory. The initial state imposes noise around imposed density values of $\rho_1 = 1.2$ and $\rho_2 = 0.6$. A value of $\tau=1$ and $\theta = 1/3$ were utilized.}
\label{fig:lbtele}
\end{figure}

To model a diffusion equation using lattice Boltzmann methods in the absence of external conservative forces, we define the tensors from Eqns. (\ref{eq:genscale0}-\ref{eq:genscale4}) as 
\begin{align}
  S &= \rho \label{eq:diffu0}\\
  j_\alpha &= 0\\
  \Phi_{\alpha\beta} &= \rho\theta\delta_{\alpha\beta}\\
  \Gamma_{\alpha\beta\gamma} &= 0\\
  \Xi_{\alpha\beta\gamma\delta} &= \frac{\rho\theta}{3}\left(\delta_{\alpha\beta}\delta_{\gamma\delta} + \delta_{\alpha\gamma}\delta_{\beta\delta} + \delta_{\alpha\delta}\delta_{\beta\gamma}\right)\label{eq:diffu4}.
\end{align}
The forcing definitions for Eqns. (\ref{eq:force0}-\ref{eq:force4}) are all set to zero. Using these definitions, we can re-derive the results from previous work by Strand et al. \cite{strandfeickert}. These moments can then be inserted into Eqn. (\ref{eq:stencil}). There are many terms which inherently vanish. With some rearranging, we arrive at a fourth-order diffusion equation in the form
\begin{align}
\partial_t\rho &= \left(\tau - \frac{1}{2}\right)\theta\nabla^2\rho +\left(\tau-\frac{1}{2}\right)\theta^2\nabla^4\rho\nonumber\\
 &- 3\left(\tau^2-\tau+\frac{1}{6}\right)\left(\tau-\frac{1}{2}\right)\theta^2\nabla^4\rho + \left(\tau^3-\frac{3}{2}\tau^2+\frac{7}{12}\tau - \frac{1}{24}\right)\theta\nabla^4\rho\nonumber\\
\partial_t\rho &= \left(\tau - \frac{1}{2}\right)\theta\nabla^2\rho + \bigg[\left(\tau^3-\frac{3}{2}\tau^2 +\frac{3}{4}\tau-\frac{1}{8}\right)\theta \nonumber\\
 &- 3\left(\tau^3-\frac{3}{2}\tau^2+\frac{2}{3}\tau-\frac{1}{12}\right)\theta + \left(\tau^3 - \frac{3}{2}\tau^2+\frac{7}{12}-\frac{1}{24}\right)\bigg]\theta\nabla^4\rho\nonumber\\
\partial_t \rho &= \left(\tau - \frac{1}{2}\right)\theta\nabla^2\rho - \left(2\tau^3\theta - \tau^3 - 3\tau^2\theta + \frac{3}{2}\tau^2 + \frac{5}{4}\tau\theta - \frac{7}{12}\tau - \frac{1}{8}\theta + \frac{1}{24}\right)\theta\nabla^4\rho.
\end{align}
Since we are utilizing a single spatial dimension, we simplified the partial differential notation in terms of $\nabla$. This gives a valid form for fourth-order corrections in the diffusive lattice Boltzmann method with a correction term
\begin{equation}
\alpha(\tau,\theta) = -\left(2\tau^3\theta - \tau^3 - 3\tau^2\theta + \frac{3}{2}\tau^2 + \frac{5}{4}\tau\theta - \frac{7}{12}\tau - \frac{1}{8}\theta + \frac{1}{24}\right).
\label{eq:diffcorr}
\end{equation}
To reach these results, the previous work made the assumption that the diffusion equation itself could be used to relate first and second order temporal derivatives to second and fourth order spatial derivatives respectively such that
\begin{align}
\partial_t \rho &= \left(\tau - \frac{1}{2}\right)\nabla^2 \rho \theta + O(\nabla^3)\label{eq:sub1}\\
\partial_t^2 \rho &= \left(\tau-\frac{1}{2}\right)^2\nabla^4 \rho \theta^2 + O(\nabla^5)\label{eq:sub2},
\end{align}
When examining Eqn. (\ref{eq:stencil}), it is immediately noticed that there are temporal derivatives of higher order which automatically included. The substitutions presented in Eqns. (\ref{eq:sub1}-\ref{eq:sub2}) are required to arrive and equation of motion which is in the form of a fourth order diffusion equation, but in the absence of these substitutions, the equation of motion takes a different form. For example, if we truncate Eqn. (\ref{eq:stencil}) to only second order using the moments defined in Eqns. (\ref{eq:diffu0}-\ref{eq:diffu4}), we have
\begin{equation}
\partial_t \rho -\left(\tau - \frac{1}{2}\right)(\partial_t^2 \rho + \partial_\alpha\partial_\beta \rho\theta\delta_{\alpha\beta}) = 0.
\label{eq:telegraph}
\end{equation}
This is not a diffusion equation, but it has a similar form to the Telegrapher's equation. In the previous analysis in \cite{strandfeickert}, Eqns. (\ref{eq:sub1}-\ref{eq:sub2}) were used to give a diffusion equation which takes the form
\begin{equation}
\partial_t \rho = D\nabla^2\rho + \alpha\nabla^4\rho
\label{eq:diffusioneq}
\end{equation}
where $D$ is the diffusion coefficient and $\alpha$ is the fourth-order correction polynomial. It is clear that the diffusion equation in Eqn. (\ref{eq:diffusioneq}) differs from the telegrapher's like equation in Eqn. (\ref{eq:telegraph}) simply due to the higher order time derivatives which are present. An interesting result of this is that when we compare the D1Q3 lattice Boltzmann simulation to central difference solutions of Eqn. (\ref{eq:telegraph}) and Eqn. (\ref{eq:diffusioneq}), we end up with identical simulation results for the telegraph like equation but differing results from the diffusion equation. Fig. \ref{fig:lbtele} shows the results over various time evolved states for a diffusion front. The symbols represent the numerical solution of Eqn. (\ref{eq:telegraph}), the X represent the central difference solution of Eqn. (\ref{eq:diffusioneq}) and the solid lines are results from the lattice Boltzmann simulation using values of $\tau = 1$ and $\theta=1/3$. The results for the telegraph like equation are seen to be identical for all times which leads to the assertion that the numerical solution to Eqn. (\ref{eq:telegraph}) is identical to the lattice Boltzmann algorithm, whereas the numerical results to the diffusion equation differ slightly over all times and do not match at the initial state. Meaning that Eqn. (\ref{eq:telegraph}) is more akin to the lattice Boltzmann algorithm that the approximated diffusion equation.

\begin{figure}
    \centering
    \includegraphics[width=\columnwidth,clip=true]{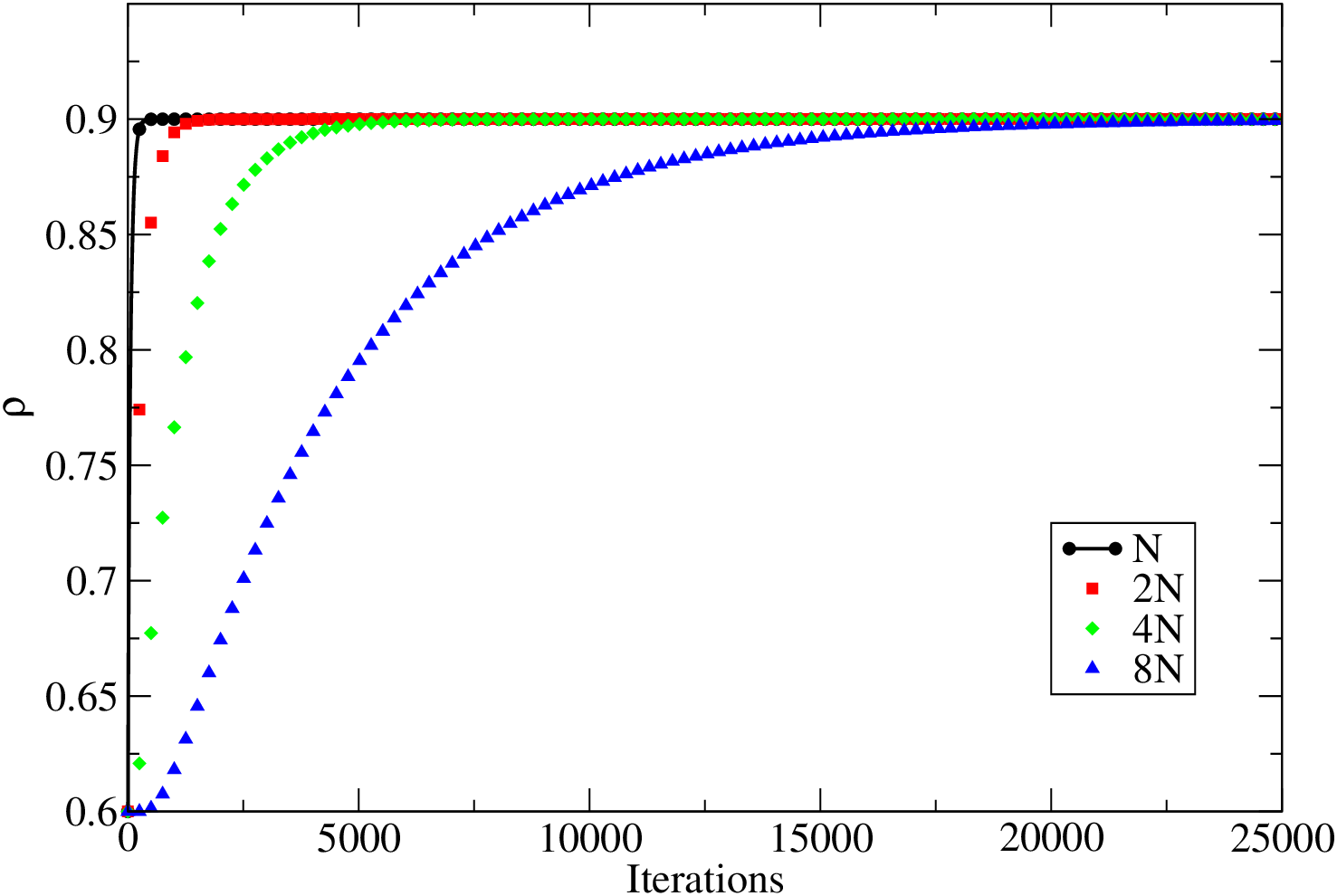}
    \caption{Plot of density over time for the fourth-order diffusion method with $N=20$ showing that as we increase the size of the lattice, the densities converge to the same value.}
    \label{fig:diffrefine}
\end{figure}

Figure \ref{fig:diffrefine} shows a lattice refinement of the fourth order method with a lattice size of $N=20$. By increasing the lattice size to various factors of $N$, we see that the method converges to the same density value.

\begin{figure}
\includegraphics[width=\columnwidth,clip=true]{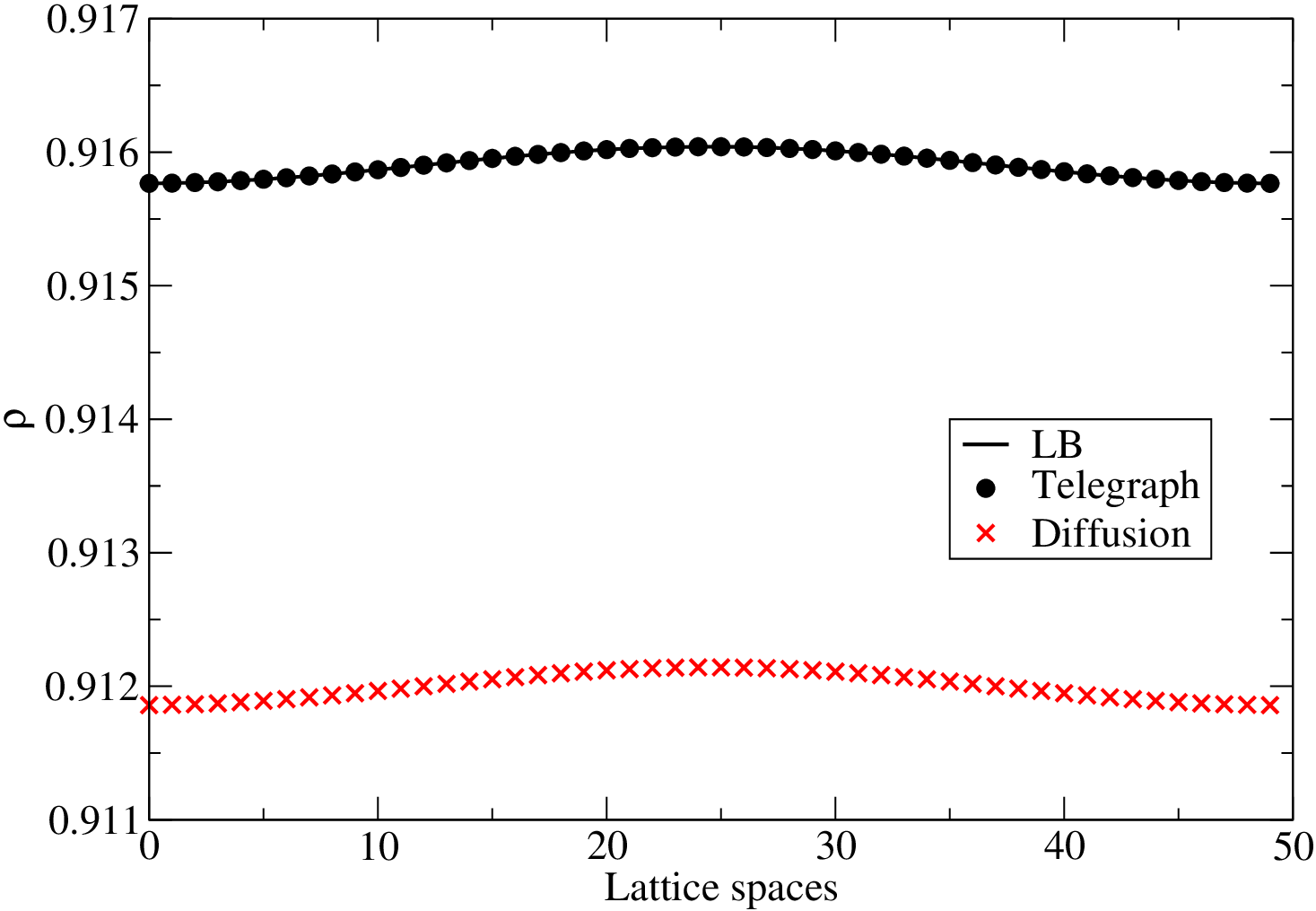}
\caption{Comparison of evolved central difference solutions of Eqn. (\ref{eq:telegraph}) (symbols), Eqn. (\ref{eq:diffusioneq}) (X), and lattice Boltzmann results (lines) after 3000 time steps. It is shown there is an approximate difference of 0.45\% at this point. Although this is small, it shows a distinct difference between the solutions of the two different equations of motion. It is also shown that the lattice Boltzmann results match the central difference solution of Eqn. (\ref{eq:telegraph}) very well.}
\label{fig:rmstele}
\end{figure}

To further demonstrate the accuracy of the central difference solution, we compare the lattice Boltzmann simulation results to the numerical results of Eqn. (\ref{eq:telegraph}) and Eqn. (\ref{eq:diffusioneq}). Figure \ref{fig:rmstele} presents results after evolving the diffusion front 3000 time steps, we observed excellent agreement between the lattice Boltzmann results and the Eqn. (\ref{eq:telegraph}). However, there is a small, but noticeable difference between the telegraph-like equation and the diffusion equation numerical solutions. Although there is a 0.45\% between the two solutions, it still implies that the telegraph-like equation models is an identical algorithm to what the lattice Boltzmann algorithm is doing. The solution to the diffusion equation still matches very well and for all practical purposes, it gives a correct result. But the underlying equation of motion could actually be Eqn. (\ref{eq:telegraph}).

\subsection{Phase separating systems}
Another example which we have employed is analyzing phase separation as governed by the Cahn-Hilliard equation. The Cahn-Hilliard system is a partial differential equation which can be derived from a flux argument using standard Fickian diffusion. The Cahn-Hilliard equation can be written 
\begin{equation}
\partial_t\rho = \nabla M \nabla \mu
\label{eq:CHEQN}
\end{equation}
where $M$ is the mobility and $\mu$ is the chemical potential. This equation shows that the density, $\rho$ will change in time as the chemical potential is modified by a Laplacian. This will cause density domains to develop until an equilibrium state is reached where $\nabla M \nabla\mu = 0$.

There are two methods which we have used to perform this analysis. Section \ref{sec:chempot} modifies the velocity moments defined for the diffusion equation to include a chemical potential and section \ref{sec:forcing} uses standard diffusion moments, but includes the chemical potential through the definition of external forcing terms.
 
\subsubsection{Chemical potential method}\label{sec:chempot}
Lattice Boltzmann methods can also be used to model systems governed by the Cahn-Hilliard equation to study phase separation \cite{foard}. For such systems, we define Eqns. (\ref{eq:genscale0}-\ref{eq:genscale4}) such that
\begin{align}
S &= \rho\\
j_\alpha &= 0\\
\Phi_{\alpha\beta} &= \mu\delta_{\alpha\beta}\\
\Gamma_{\alpha\beta\gamma} &= 0\\
\Xi_{\alpha\beta\gamma\delta} &= \frac{\mu}{3}(\delta_{\alpha\beta}\delta_{\gamma\delta} + \delta_{\alpha\gamma}\delta_{\beta\delta} + \delta_{\alpha\delta}\delta_{\beta\gamma}).
\end{align}
Once again, we set the definitions for the forcing moments to be equal to zero. Inserting these moments into Eqn. (\ref{eq:stencil}) will allow us to define a fourth order Cahn-Hilliard equation. In the initial substitution, we will be left with higher order temporal derivatives:
\begin{align}
\partial_t \rho -\left(\tau - \frac{1}{2}\right)&(\partial_t^2\rho + \nabla^2\mu) + 3\left(\tau^2 - \tau + \frac{1}{6}\right)\partial_t\nabla^2\mu\nonumber\\ 
&- \left(\tau^3 - \frac{3}{2} \tau^2 + \frac{7}{12}\tau - \frac{1}{24}\right)\nabla^4\mu = 0\\.
\end{align}
In order to deal with these, we must define substitutions similar to those in Eqns. (\ref{eq:sub1}-\ref{eq:sub2}) from the diffusive case. For the Cahn-Hilliard equation, we naively define these as
\begin{equation}
\partial_t\rho = \left(\tau - \frac{1}{2}\right)\nabla^2\mu
\end{equation}
\begin{equation}
\partial_t^2 \rho = \left(\tau - \frac{1}{2}\right)^2 \nabla^4\mu.
\end{equation}
With these definitions, we can derive a fourth-order Cahn-Hilliard equation. Following the algebra, we are left with
\begin{align}
\partial_t \rho = \left(\tau - \frac{1}{2}\right) \nabla^2 \mu &+ \bigg[\left(\tau - \frac{1}{2}\right)^3 -3\left(\tau - \frac{1}{2}\right)\left(\tau^2 - \tau + \frac{1}{6}\right)\nonumber\\ 
&+ \left(\tau^3 - \frac{3}{2}\tau^2 + \frac{7}{12}\tau - \frac{1}{24}\right)\bigg]\nabla^4 \mu\nonumber\\
\end{align}
\begin{align}
\partial_t \rho = \left(\tau - \frac{1}{2}\right)\nabla^2 \mu +\left(-\tau^3 + \frac{3}{2}\tau^2 - \frac{2}{3}\tau +\frac{1}{12}\right)\nabla^4\mu
\label{eq:fdchempot}
\end{align}
with a correction term defined as
\begin{equation}
\alpha(\tau) = \left(-\tau^3 + \frac{3}{2}\tau^2 - \frac{2}{3}\tau +\frac{1}{12}\right).
\end{equation}
In this case, the correction terms is only a function of $\tau$ where the correction term in Eqn. (\ref{eq:diffcorr}) is a function of both $\tau$ and $\theta$. This arises from the simple fact that $\theta$ is absent from the moments defined for Cahn-Hilliard systems. The moment for the diffusive systems have the lattice temperature $\theta$ inherently included.

\begin{figure}
\includegraphics[width=\columnwidth,clip=true]{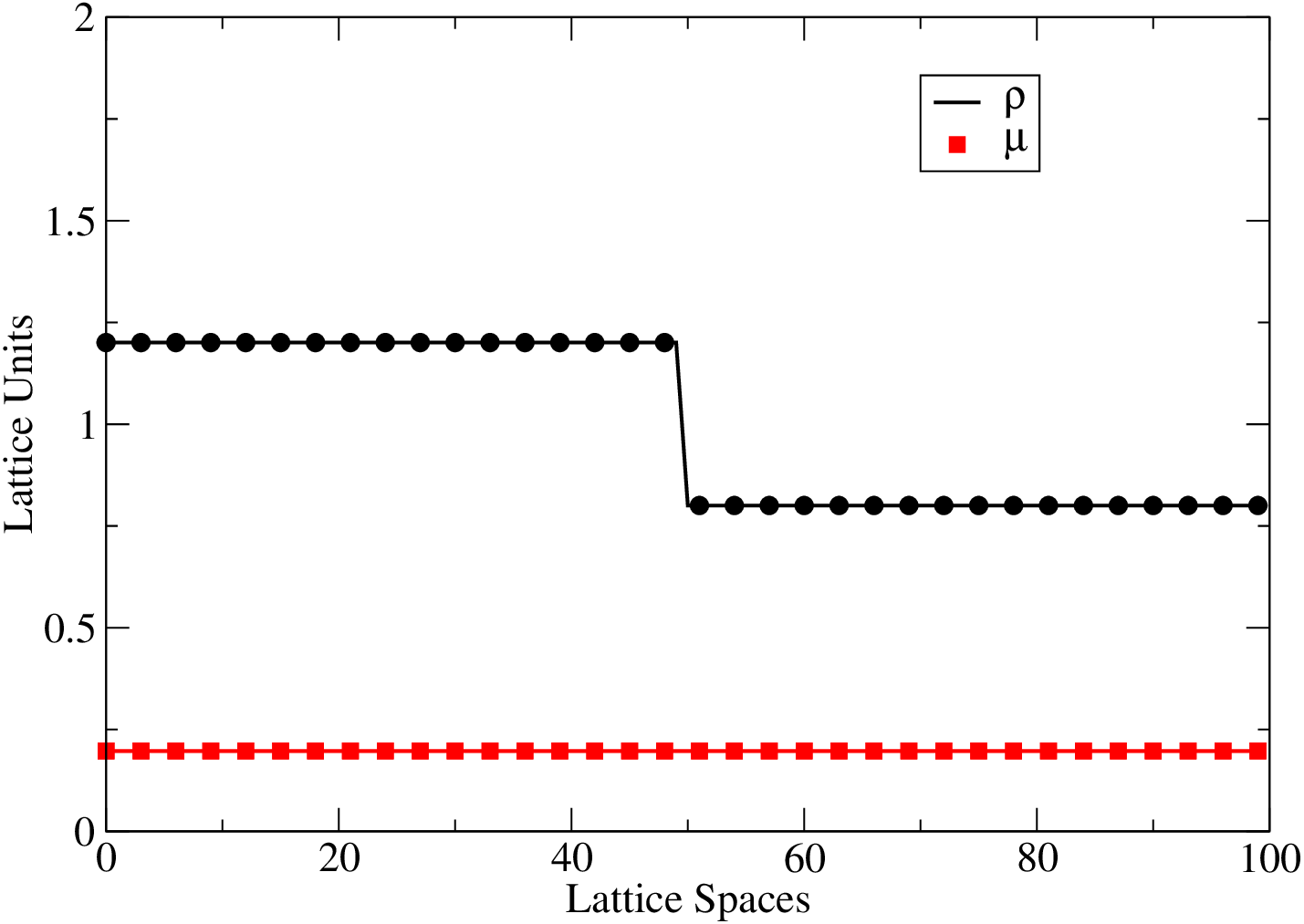}
\caption{Comparison of lattice Boltzmann and theoretical results for $\rho$ and $\mu$ using a finite difference solution of Eqn. (\ref{eq:fdchempot}) for the theory results with $\theta_1 = 0.16667$, $\theta_2 = 0.25466$, $\rho_1 = 1.2$, $\rho_2 = 0.8$, and $M=0.5$. We see good agreement for both $\rho$ and $\mu$ across the entire lattice. }
\label{fig:rhomu}
\end{figure}

To justify the chemical potential method for a Cahn-Hilliard system, we introduce a simple system with two domains that each contain distinct temperatures, $\theta_1$ and $\theta_2$. With these temperature domains, we introduce a simple free energy which takes the form
\begin{equation}
A(x,t)=\theta_{1,2}\rho\ln\rho
\label{eq:freenrg}
\end{equation}
where $\theta_{1,2}$ refers to the temperature of a specific domain. The chemical potential follows by the thermodynamic relation
\begin{equation}
\mu(x,t) = \frac{\partial A}{\partial \rho} = \theta_{1,2} (1+\ln\rho).
\label{eq:chempot}
\end{equation}
To relate $\theta_1$ and $\theta_2$, we make the ansatz that 
\begin{equation}
\theta_1 (1+\ln\rho_1) = \theta_2 (1+\ln\rho_2)
\end{equation}
where $\rho_1$ and $\rho_2$ are initial densities for the corresponding domains. With this, we can solve for $\theta_2$ in terms of $\theta_1$ and the initial densities such that
\begin{equation}
\theta_2 = \theta_1 \frac{1+\ln\rho_1}{1+\ln\rho_2}.
\end{equation}
With this system, we can employ the lattice Boltzmann model from previously in the section. Figure \ref{fig:rhomu} compares the lattice Boltzmann results to theory for both $\rho$ and $\mu$ where the symbols are the simulation results and the solid lines are theory. For this system, we have chosen $\theta_1 = 0.16667$ which gives $\theta_2 = 0.25466$. For our initial densities, we have chosen a random distribution around $\rho_1 = 1.2$ and $\rho_2 = 0.8$. For the Cahn-Hilliard system, we have introduced a spatial and time dependent relaxation time $\tau(x,t)$ which is related through a constant mobility like parameter $M$. To find this relaxation time, we use the relation
\begin{equation}
\tau(x,t) = M\rho + \frac{1}{2}.
\end{equation}
These results have a value of $M=0.5$. We see very good agreement between the theoretical predictions and simulations for density and chemical potential. Since the system is allowed to equilibrate, we see a constant constant chemical potential over the entire lattice. The theoretical values represented on this plot come from a finite difference solution to Eqn. (\ref{eq:fdchempot}). However, although there is initial noise which as been applied in each domain, this solution can itself be considered a stationary state solution. Future work includes studying a system by observing dynamics the dynamics in the frame of fluctuating lattice Boltzmann methods. 

\begin{figure}
    \centering
    \includegraphics[width=\columnwidth,clip=true]{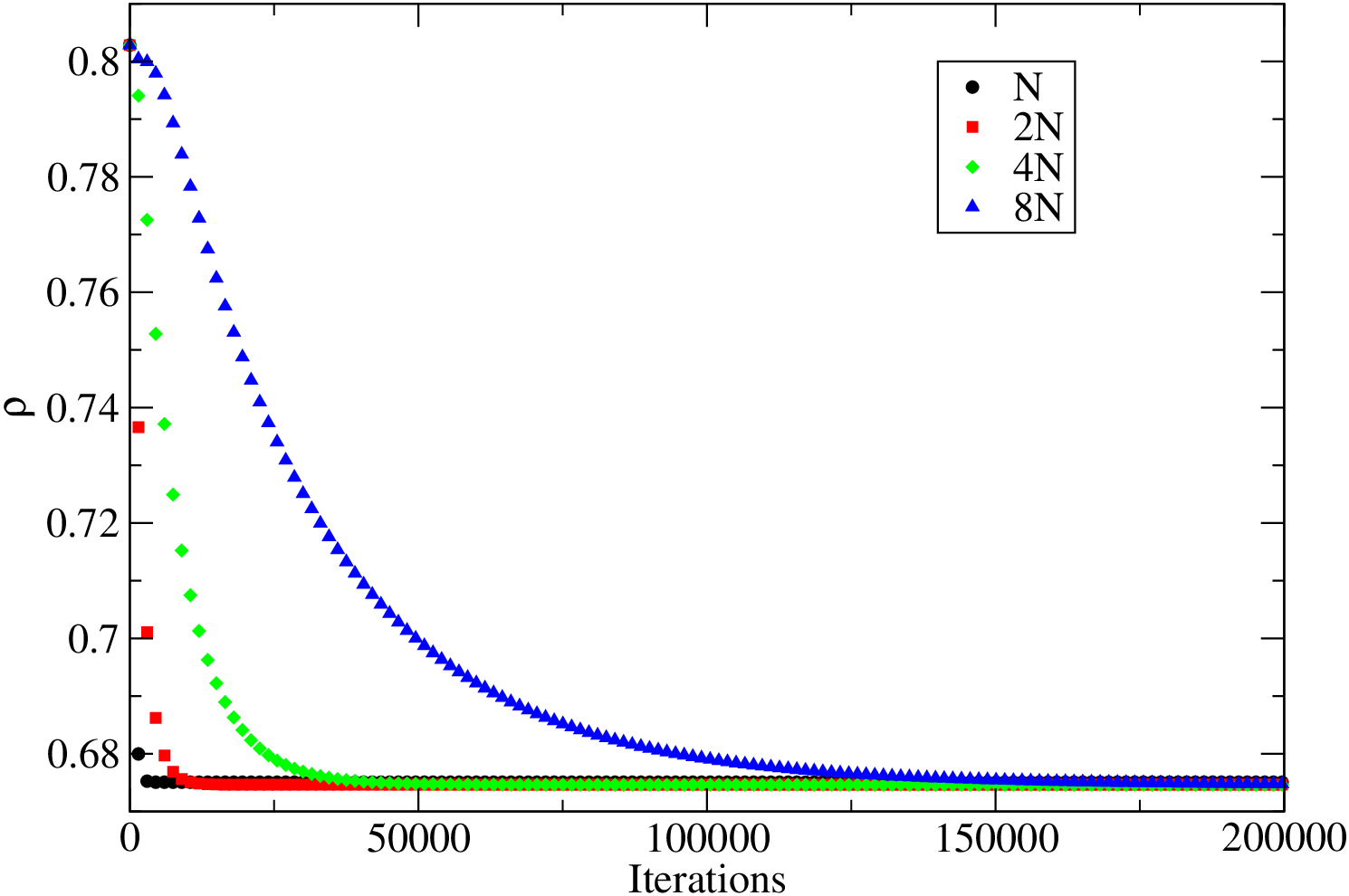}
    \caption{Plot of density over time for the fourth-order chemical potential method with $N=20$ showing that as we increase the size of the lattice, the densities converge to the same value.}
    \label{fig:murefine}
\end{figure}

To further illustrate the validty of the fourth order method, Figure \ref{fig:murefine} shows a lattice refinement of the fourth order method with a lattice size of $N=20$ and increasing the lattice size by factors of $N$. As we increase the size of the lattice, we see the densities of each size converge to the same value.

\subsubsection{Forcing method}\label{sec:forcing}
Here we present an identical system as presented in the previous section with the free energy and chemical potential represented by Eqns. (\ref{eq:freenrg} - \ref{eq:chempot}) respectively. However, in this case we employ a forcing method as opposed to the chemical potential method previously presented. We choose the same equilibrium moments defined in Eqns. (\ref{eq:diffu0}-\ref{eq:diffu4}), but we must also define forcing moments from Eqns. (\ref{eq:force0} - \ref{eq:force4}). In order to do this, we propose a forcing term $F_\alpha$ which is governed by a gradient of the non-ideal part of the chemical potential. We define
\begin{equation}
F_\alpha = \left(1-\frac{1}{2\tau}\right)\rho\partial_\alpha\mu^{nid}(\bm{x},t)
\end{equation}
where $\mu^{nid}(\bm{x},t)$ is the non-ideal part of the chemical potential. The prefactor on this equation is the same prefactor which was presented by Guo et al.\cite{guoetal} and He et al.\cite{heetal} in their derivations for lattice Boltzmann schemes with external forces. We can write a full chemical potential in terms of the ideal part, $\mu^{id}(\bm{x},t) = \theta\ln\rho$, and the non-ideal part such that
\begin{equation}
\mu(\bm{x},t) = \mu^{id}(\bm{x},t) + \mu^{nid}(\bm{x},t).
\end{equation}
Now, by solving this full chemical potential for $\mu^{nid}(x,t)$, taking a gradient of the non-ideal part, and multiplying by a factor of $\rho$, we are left with a forcing definition in the form
\begin{equation}
F_\alpha = \left(1-\frac{1}{2\tau}\right)(\rho\partial_\alpha\mu - \theta\partial_\alpha\rho)
\end{equation}
We can now use this definition for $F_\alpha$ as our guide to defining the forcing moments required for this model. Since we are utilizing a one dimensional model, we will drop all indices in reference to spatial dimensions and use $\nabla$ in place of partial derivatives for simplicity.
\begin{align}
F_\alpha &= \left(1-\frac{1}{2\tau}\right)(\rho\nabla\mu - \theta\nabla\rho) \label{eq:forced0}\\
\Psi_{\alpha\beta} &= \Psi\\
\Delta_{\alpha\beta\gamma} &= F_\alpha = \left(1-\frac{1}{2\tau}\right)(\rho\nabla\mu - \theta\nabla\rho)\\
Z_{\alpha\beta\gamma\delta} &= \Psi. \label{eq:forced4}
\end{align}
Note that the first and third moment and second fourth moment are identical. We choose this as to not change the dynamics of the system by performing a higher order analysis. Since we are just introducing a correction term, we expect no change to the dynamics, thus each higher order moment introduced, must reflect that which was utilized in the second order implementation. It is also important to point out that we have not yet defined a form for the second moment $\Psi$. We will find the proper form of this moment by substituting these moments into our expansion and find the proper form which then reproduces the Cahn-Hilliard equation. 

For simplicity of the current derivation, we are only going to examine the stationary state solution of the Cahn-Hilliard equation such that
\begin{equation}
    \nabla M \nabla \mu = 0
    \label{eq:stationarych}
\end{equation}
where we examine a stationary equilibrium profile which has all temporal derivatives equal to zero and is produced through the same moments presented Eqns. (\ref{eq:forced0}-\ref{eq:forced4}). This is done in a similar manner as Wagner \cite{wagner3} where he used stationary profiles to examine higher order lattice Boltzmann models for the Navier-Stokes equations. One important caveat to note is that this analysis only addresses the equilibrium behavior of the system and does not consider the overall dynamics which could be seen in true phase separating systems governed by the Cahn-Hilliard equation. Full dynamics of the fourth-order Cahn-Hilliard lattice Boltzmann will be addressed in future work.

By substituting these definitions into Eqn. (\ref{eq:stencil}) with at temporal derivatives set to zero, we initially arrive at
\begin{align}
    -\lambda_2(\tau)\left[\nabla(\rho\nabla\mu - \theta\nabla\rho) + \theta\nabla^2\rho - \tau\nabla^2\Psi\right] - &\lambda_2(\tau)\lambda_3(\tau)\nabla^3(\rho\nabla\mu - \theta\nabla\rho) -\nonumber\\
    &\lambda_4(\tau)(\theta\nabla^4\rho -\tau \nabla^4\Psi) = 0
    \label{eq:stationary1}
\end{align}
where we have used the Bernoulli polynomial notation from Eqns. (\ref{eq:bern1}-\ref{eq:bern4}). Note that we are left with a fourth-order term for $\Psi$ in this previous equation. If we were to perform a third order analysis on these moments, we would see that $\Psi = O(\partial^2)$. From this, we make the assumption that it is of the same order for the fourth-order case. With this, we have $\nabla^4\Psi = O(\partial^6)$ so we are able to neglect the final $\nabla^4\Psi$ term from Eqn. (\ref{eq:stationary1}). A full derivation for the third-order form of $\Psi$ is written in \ref{sec:append2}.

Now, we are able to carry out the next steps of the derivation by expanding out derivatives in the first-order forcing term and simplifying:
\begin{equation}
     -\lambda_2(\tau)\nabla(\rho\nabla\mu) + \lambda_2(\tau)\tau\nabla^2\Psi - \lambda_2(\tau)\lambda_3(\tau)\nabla^3(\rho\nabla\mu - \theta\nabla\rho) -\lambda_4(\tau)\theta\nabla^4\rho = 0.
    \label{eq:stationary2}
\end{equation}
Examining this equation, we see the first term has the form of the stationary Cahn-Hilliard equation where in this case we have the mobility such that $M = \lambda_2(\tau)\rho$. With this, we recognize that for the equilibrium solution to hold, we must have 
\begin{equation}
    \lambda_2(\tau)\nabla(\rho\nabla\mu) = 0.
\end{equation}
Since this is the case, we know that we also must have 
\begin{equation}
    \lambda_2(\tau)\tau\nabla^2\Psi -\lambda_2(\tau)\lambda_3(\tau)\nabla^3(\rho\nabla\mu - \theta\nabla\rho) -\lambda_4(\tau)\theta\nabla^4\rho = 0.
\end{equation}
Now with this, we can simply solve for the amending function $\Psi$ and will will have our fourth-order correction to this illustrative equilibrium example. Doing this, our amending function has the form
\begin{equation}
    \Psi = \frac{\lambda_3(\tau)}{\tau}\nabla(\rho\nabla\mu -\theta\nabla\rho) + \frac{\lambda_4(\tau)\theta}{\lambda_2(\tau)\tau}\nabla^2\rho.
\end{equation}

\begin{figure}
    \centering
    \includegraphics[width=\columnwidth,clip=true]{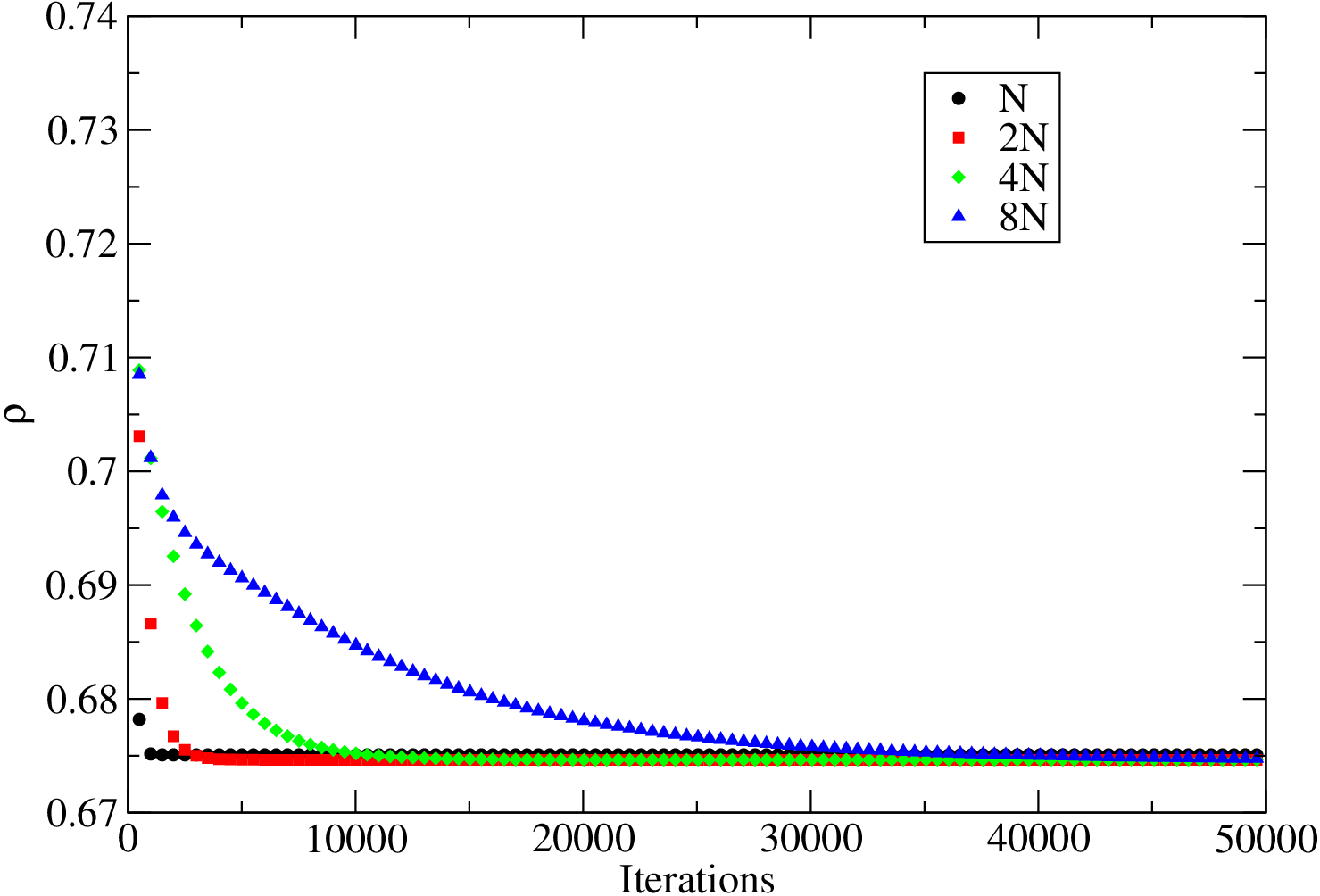}
    \caption{Plot of density over time for the fourth-order forcing method with $N=20$ showing that as we increase the size of the lattice, the densities converge to the same value.}
    \label{fig:forcerefine}
\end{figure}

First, we demonstrate a lattice refinement of the fourth-order forcing method. Figure \ref{fig:forcerefine} shows density over time with a lattice size of $N=20$. As we increase the lattice size by various factors of $N$, we see that the method converges to the same value.

\begin{figure}
\includegraphics[width=\columnwidth,clip=true]{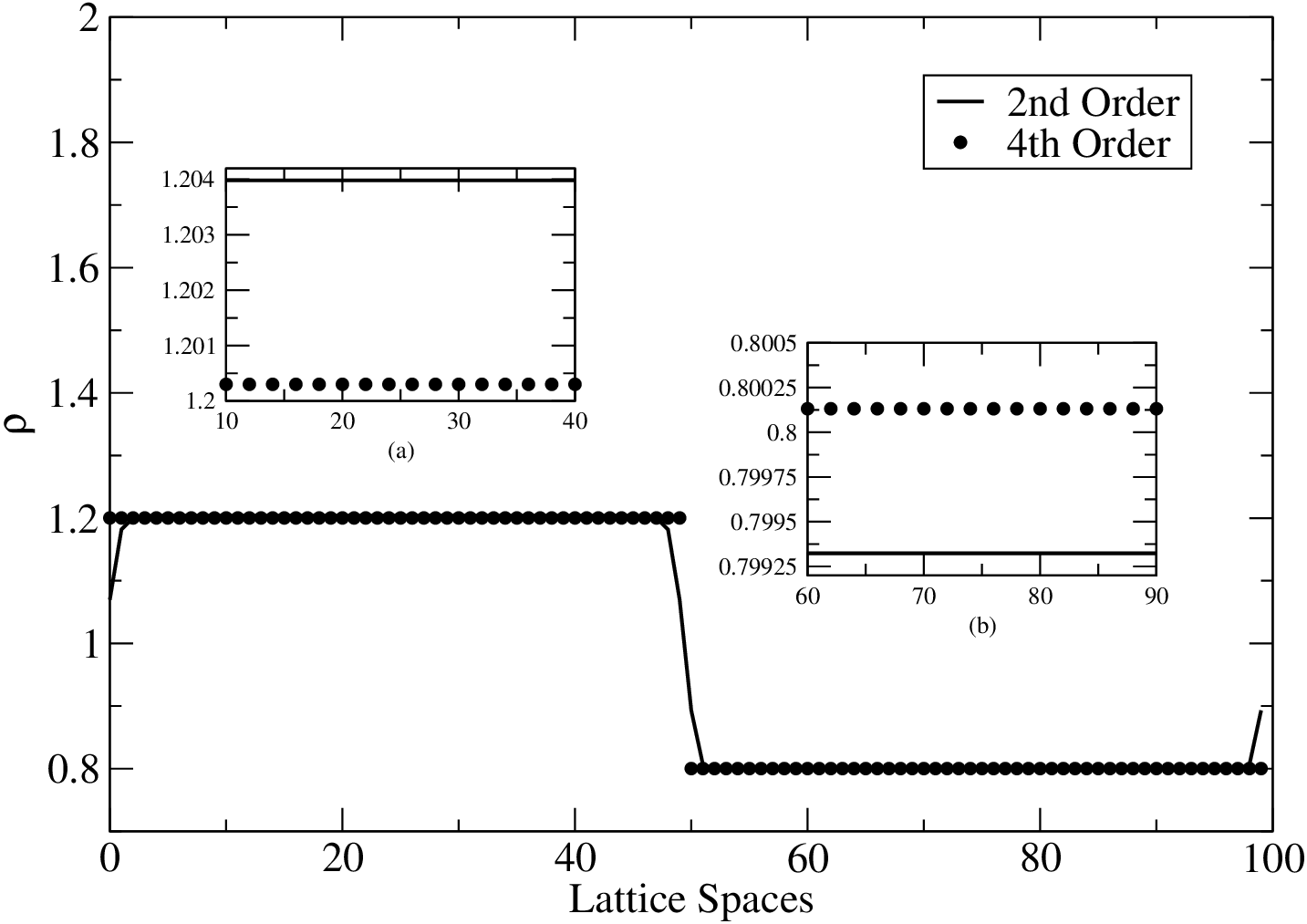}
\caption{Comparison of second (solid lines) and fourth (symbols) order densities with $\theta_1 = 0.16667$, $\theta_2 = 0.25466$, $\rho_1 = 1.2$, $\rho_2 = 0.8$, and $M=0.5$. On the scale of the entire lattice, we see that the second and fourth order methods are very similar. Inset (a) zooms in on the left domain equilibrium values. We see that the second order method gives $\rho_1 \approx 1.204$ and the fourth order method gives $\rho_1 \approx 1.2005$. Inset (b) zooms in on the right domain equilibrium values. This domain has a second order method value of $\rho_2 \approx 0.79935$ and the fourth order gives $\rho_2 \approx 0.800125$. Each domain shows slight improvement from the fourth order method over the second order method. }
\label{fig:rhoforce}
\end{figure}

Since we have prepared an identical system as in the previous section, we expect similar results which were presented for the chemical potential model. We have once again used the parameters $\theta_1=0.16667$, $\theta_2 = 0.25466$, $\rho_1 = 1.2$, $\rho_2 = 0.8$, and $M=0.5$. Using this system, we have run simulations using a traditional second order forcing method as well as the new fourth order method. Figure \ref{fig:rhoforce} compares the second (solid lines) and fourth order densities (symbols) which result from the method. Immediately we see that the results are on the same order as we saw in section \ref{sec:chempot}. For this reason, we have not included the theory lines on this plot. However, at this big level, we see that there does not seem to be a distinct difference between the second and fourth order methods. Inset (a) zooms in closely on the left domain and a noticeable difference is noted. The second order method gives an equilibrium density in this domain of $\rho_1 \approx 1.204$ while the fourth order gives a result of $\rho_1 \approx 1.2005$. The right domain is represented by inset (b) which gives a second order result of $\rho_2 \approx 0.79935$ and a fourth order result of $\rho_2 \approx 0.800125$.

\begin{figure}
\includegraphics[width=\columnwidth,clip=true]{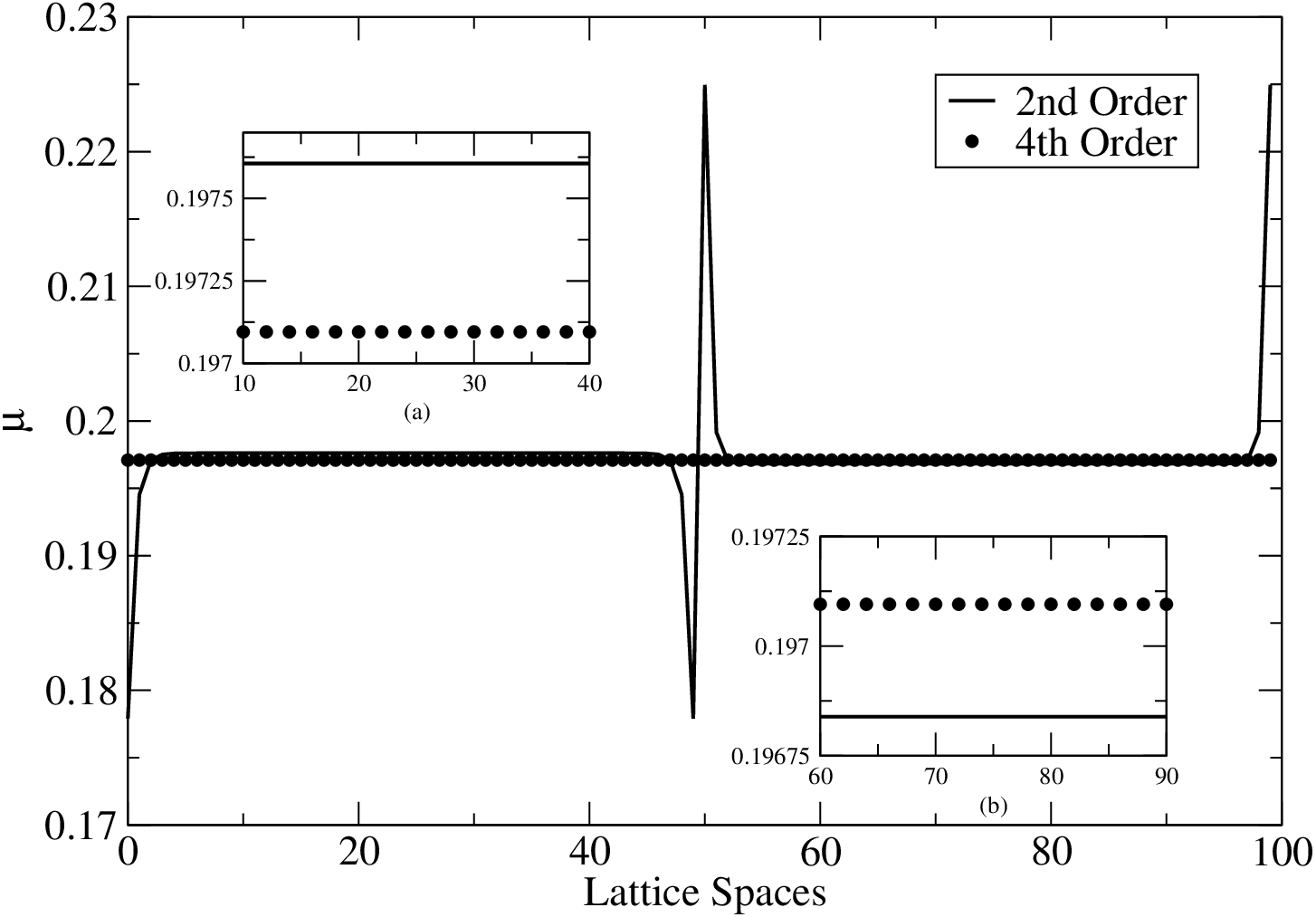}
\caption{Comparison of second (solid lines) and fourth (symbols) order chemical potentials with $\theta_1 = 0.16667$, $\theta_2 = 0.25466$, $\rho_1 = 1.2$, $\rho_2 = 0.8$, and $M=0.5$. For the full scale of the lattice, we see overall agreement between the second and fourth order methods. However, we see an asymptotic line at the location of the lattice interface for the second order method which is corrected by the fourth order method. Inset (a) zooms in on the left domain where the second order method takes a value of $\mu_1 \approx 0.1976$ and the fourth order giving $\mu_1 \approx 0.1971$. Inset (b) zooms in on the right domain and the second order method gives $\mu_2 \approx 0.1968$ and the fourth order takes a value $\mu_2 \approx 0.1971$. In equilibrium, we see a very small difference between the chemical potentials where we would expect them to be a constant value at equilibrium. However, this difference is small enough that it can be neglected. The major advantage of the fourth order method is that it corrects the asymptotic divergence at the interface seen in the second order method.}
\label{fig:muforce}
\end{figure}

In figure \ref{fig:muforce}, we examine the difference between the second and fourth order chemical potentials. Once again, we see agreement between the second and fourth order and the theoretical result show in figure \ref{fig:rhomu}. We see a near constant chemical potential, however the major difference here is that at there is an asymptotic line at the interface for the second order method. In this case, the fourth order method gives a drastic improvement in that the asymptotic behavior at the interfaces is completely removed, thus giving rise to true equilibrium behavior. Inset (a) zooms in on the left side domain and we see $\mu_1 \approx 0.1976$ for second order and $\mu_1 \approx 0.1915$ for fourth order. Inset (b) looks at the right domain which gives $\mu_2 \approx 0.1968$ for second order and $\mu_2 \approx 0.1971$ for fourth order. As seen in the density representation from figure \ref{fig:rhoforce}, we see improvement to the accuracy with the inclusion of fourth order terms. When employing the forcing method for the Cahn-Hilliard system, the fourth-order method offers a large improvement over the second-order.

\section{Conclusions}
In this manuscript, we have proposed a moment independent expansion for examining the lattice Boltzmann method to the fourth-order. This has been done by taking an expansion of the lattice Boltzmann equation up to fourth order and taking its hydrodynamic limit. In the hydrodynamic limit, we have been able to show how to apply specific velocity moments to the expansion to find the fourth-order equations of motion and also define correction terms to the method. We then applied this moment independent method to various different systems and verified the fourth order method against a traditional second order method. First, we re-derived results previously derived for a diffusive system using this new moment independent expansion. This method utilizes a substitution for replacing temporal derivatives with spatial derivatives was discussed in Eqns. (\ref{eq:sub1}-\ref{eq:sub2}). It was shown that the actual lattice Boltzmann expansion gives rise to higher order temporal derivatives which may be of some importance. The diffusive lattice Boltzmann method was compared to a finite difference solution of Eqn. (\ref{eq:telegraph}) for a diffusion front. We observed that the lattice Boltzmann algorithm matches the numerical solution almost identically. We then presented two methods for modeling phase separating systems. The first method is a chemical potential based model which includes no external forces that gives rise to a Cahn-Hilliard equation. We examined the fourth order method compared to the second order where the fourth order gave a slight improvement over the second order. We also observed a constant chemical potential across the entire lattice in equilibrium for both second and fourth order methods. Secondly, we used external forces in addition to the traditional diffusive lattice Boltzmann moments. In the case of the forced system, we observed that the second order method did phase separate, however the chemical potential was asymptotic at the interface. In implementing the fourth order correction, we saw improvement over the second order method in terms of accuracy and by removing the asymptotic line at the interface giving a constant chemical potential. The constant chemical potential across the lattices gives true equilibrium behavior which is not seen at second order. For the case of the forcing method, the fourth order gives drastic improvement to the second order method.

We have seen in the diffusive system and the chemical potential method for the phase separating system that the fourth order method only gave minor improvement over the traditional second order methods. Since this is the case, it is unnecessary to perform a fourth order analysis for these systems. However, in the case of the forcing method for phase separating systems. We saw large improvement in the sheer fact that it gives a constant chemical potential at equilibrium rather than the asymptotic chemical potentials at equilibrium. In this case, we would certainly want to use the fourth order method. In closing, it is important to state that the usefulness of this expansion is highly dependent on the system which is being simulated and the method in which the physical system is modeling as certain cases offer large improvement whereas other only offer minor improvements.

\appendix

\section{Extension to arbitrary order}\label{sec:append}
Here we introduce the method of generalizing the fourth order expansion to arbitrary order. Beginning with Eqn. (\ref{eq:fi0pde}), we first recognize that this can be written as a series. We can then rewrite this equation as
\begin{equation}
\sum_{m=1}^4 \lambda_m(\tau) \frac{(\partial_t +v_{i\alpha}\partial_\alpha)^m}{m!}(f_i^0 + \tau F_i) + O(\partial^5) = \Omega_i
\end{equation}
where $\lambda_m(\tau)$ are the Bernoulli polynomials for each specific order. We can generalize this series to arbitrary order by extending the limits on the sum such that
\begin{equation}
\sum_{m=1}^n \lambda_m(\tau) \frac{(\partial_t +v_{i\alpha}\partial_\alpha)^m}{m!}(f_i^0 + \tau F_i) + O(\partial^{n+1}) = \Omega_i
\end{equation}
where $n$ is the desired order of the expansion. For simplicity sake, we define the sum on the left hand side as
\begin{equation}
\chi^n \equiv \sum_{m=1}^n \lambda_m(\tau) \frac{(\partial_t +v_{i\alpha}\partial_\alpha)^m}{m!}(f_i^0 + \tau F_i) + O(\partial^{n+1}).
\end{equation}
Now, to attain the equations of motion, we then sum both sides over all $i$ and we are left with
\begin{equation}
\sum_i \chi^n = \sum_i \Omega_i.
\label{eq:almost}
\end{equation}
This simple and concise form is valid for systems with a single conserved quantity, but this can be generalized further to account for systems which require more than one conserved quantity. In general, to acquire the equations of motion for additional conserved quantities, we multiply Eqn. (\ref{eq:fi0pde}) by powers of $v_{i\alpha}$ which correspond to the moments in Eqns. (\ref{eq:genscale0}-\ref{eq:genscale4}). We can define a product over these velocities in the form
\begin{equation}
\eta_c = \prod_{j=0}^c v_{j\alpha}
\end{equation}
with $c$ representing the order of the moment which is required for any desired conserved quantity. Combining these products with Eqn. (\ref{eq:almost}), we then arrive at
\begin{equation}
\sum_i \eta_c \chi^n = \sum_i \eta_c \Omega_i.
\end{equation}
This equation is a simple mathematical statement representing the hydrodynamic limit of any lattice Boltzmann method to arbitrary order and conserved quantity.

\section{Third-order stationary derivation for forcing method}\label{sec:append2}
Here we will briefly derive the third-order amending function $\Psi$ for an equilibrium profile of the Cahn-Hilliard equation such that 
\begin{equation}
    \nabla(M\nabla\mu) = 0.
\end{equation} 
Using Eqn. (\ref{eq:stencil}) up to third order with the forcing moments from Eqns. (\ref{eq:forced0}-\ref{eq:forced4}), we are left with
\begin{equation}
    -\lambda_2(\tau)\left[\nabla(\rho\nabla\mu - \theta\nabla\rho) + \theta\nabla^2\rho - \tau\nabla^2\Psi\right] - \lambda_2(\tau)\lambda_3(\tau)\nabla^3(\rho\nabla\mu - \theta\nabla\rho) = 0.
\end{equation}
Note that we have set all temporal derivatives in the expansion to zero in the same manner as in section \ref{sec:forcing}. Taking the gradient of the first order forcing term and simplifying gives
\begin{equation}
    -\lambda_2(\tau)\nabla(\rho\nabla\mu) + \lambda_2(\tau)\tau\nabla^2\Psi- \lambda_2(\tau)\lambda_3(\tau)\nabla^3(\rho\nabla\mu - \theta\nabla\rho) = 0.
\end{equation}
Recognizing that the first term in the previous equation takes the form of the Cahn-Hilliard equation with $M=\lambda_2(\tau)\rho$, we know that
\begin{equation}
\lambda_2(\tau)\nabla(\rho\nabla\mu) = 0.    
\end{equation}
Because of this, we must have 
\begin{equation}
\lambda_2(\tau)\tau\nabla^2\Psi- \lambda_2(\tau)\lambda_3(\tau)\nabla^3(\rho\nabla\mu - \theta\nabla\rho) = 0.
\end{equation}
Solving this equation for $\Psi$ gives
\begin{equation}
    \Psi = \frac{\lambda_3(\tau)}{\tau}\nabla(\rho\nabla\mu - \theta\nabla\rho).
\end{equation}
With this solution, we deduce that $\Psi = O(\partial^2)$.

\newpage
\bibliographystyle{elsarticle-num-names}
\biboptions{square,numbers,sort&compress}
\bibliography{fourth}

\end{document}